\documentclass[a4paper,12pt]{article}
\usepackage{epsfig,multicol}

\newcommand{\beq}{\begin{equation}}
\newcommand{\eeq}{\end{equation}}
\newcommand{\bra}{\begin{array}}
\newcommand{\era}{\end{array}}

\newcommand{\al}{\alpha}

\newcommand{\de}{\delta}

\usepackage{graphicx}

\usepackage{latexsym}

\author{Jamila Douari\footnote{jdouari@hotmail.com}\\ \\
\small\it The Abdus Salam International Centre for Theoretical Physics,\\
\small\it High Energy Section Trieste, Italy\rm }
\title{The Curved Space is The Electrified Flat Space}
\begin{document}
\maketitle \vspace*{0.5cm} \maketitle \vspace*{0.5cm} PACS: 11.25.-w; 11.25.Uv; 11.15.Kc; 11.27.qd \vskip1cm
Keywords: Branes, Strings, Dyons, Perturbations, Supergravity. \vspace*{1.5cm}
\section*{Abstract}
\hspace{.3in}The responsibility of the electric field $E$ in the modification of the nature of the space is proved. We investigate the way the fundamental strings are related to supergravity background of D5-branes; i.e. once the endpoints of the D-strings are electrified the flat space becomes curved. We study the electrified relative and overall transverse perturbations of fuzzy funnel solutions of intersecting $(N,N_f)$-strings and D5-branes in flat and supergravity backgrounds respectively. As result the perturbations have a discontinuity which corresponds to a zero phase shift realizing Polchinski's open string Neumann boundary condition. And once the electric field $E$ is turned on in flat space these perturbations decrease and when $E$ is close to the critical value $\frac{1}{\lambda}$ the perturbations disappear forever and the string coupling becomes strong. At this stage the space is considered curved and the electric field is responsible for this effect. This phenomena is also enhanced by the behavior of the potential $V$ associated to the perturbations $\Phi$ on the funnel solutions under the influence of the electric field. The potential goes too fast to $-\infty$ when $E$ goes to the critical value $\frac{1}{\lambda}$ in flat space which looks like a kink to increase the velocity for $\Phi$ to disappear. But in curved space and close to the intersecting point we do not find any perturbation for all $E$ and there is no effect of $E$ on $V$ and this is a sign to the absence of the perturbation effects in supergravity background. This clarifies the existence of a relation between the electric field and the supergravity background.

\section{Introduction}
\hspace{.3in}The present work proves the fact that the flat space becomes curved because of the presence of the electric field. We use the non-Abelian Dirac-Born-Infeld (DBI) effective action for this study. Many results using this action have dealt with brane intersections and polarization \cite{BI,InterBran1,fun,InterBran2,9911136,supergravComp}. The study of brane intersections has given a realization of non-commutative geometry in the form of so-called fuzzy funnels \cite{dual,fuzfun,d1d3,TW,d1d3duality,d1d2d3d5duality,d1d3d5}. In the context of time dependence in string theory from the effective D-brane action, we expect that the hyperplanes can fluctuate in shape and position as dynamical objects.

We deal with the branes intersection problem of $(N,N_f)$-strings with D5-branes in flat and curved spaces by treating the relative and overall transverse perturbations. And it will be devoted to extend the research begun in \cite{d1d3,d1d2d3d5duality,d1d3d5}. The duality of intersecting D1-D3 branes in the low energy effective theory in the presence of electric field is found to be broken in \cite{d1d3duality} but the duality of intersecting D1-D5 branes discussed in \cite{d1d2d3d5duality} is unbroken in the same theory with the electric field is switched on which allows us to be more interested by the study of the intersecting D1-D5 branes.

We observe, in section 2, that the most lowest energy is gotten as the electric field $E$ is approximately its critical value $\frac{1}{\lambda}$ ($\lambda=2\pi \ell_s ^2$ and $\ell_s$ the string length) and also as $E$ is going to $\frac{1}{\lambda}$ the physical radius is going to the highest value and then D5-brane is getting bulky.

The analysis we give in sections 3 and 4 proves that the perturbations have a discontinuity which corresponds to a zero phase shift and then the string is Polchinski's open string obeying Neumann boundary condition. Hence the endpoints lie on the hyperplane are still free to move in.

We also look for more effects of $E$ on the perturbations and the associated potentials. The behavior of the perturbations in both backgrounds is as follows: in flat space (section 3), the perturbations are disappearing because of the presence of $E$ and when $E\approx \frac{1}{\lambda}$ we end by no perturbation and our system is stable; and in curved space (section 4) we did not get any perturbation for all $E$ which means the presence of the supergravity does not allow any perturbation to appear in the same way that $E$ does in flat space. 

The effect of $E$ on the potentials associated to the perturbations in flat and curved spaces is the following: the potential is going down too fast to a very low amplitude minima ($-\infty$) in flat space as $E$ is going to its maxima, this is interpreted as inducing an increase in the velocity of the perturbation to disappear; and in curved space the effect of $E$ on the potential is absent.

The comparison of the flat and curved cases leads us to say if $E$ or supergravity is present then the perturbations should be absent. This looks like $E$ affects the flat background of D5-brane and transformed it to supergravity background where the objects are stable. Consequently, we can think of $E$ and supergravity as dual.

It's known that in curved space the string coupling $g_s$ is strong. And from our study the electric field $E$ is fixed in terms of $g_s$ by the relation $E=\frac{1}{\lambda}(1+(\frac{N}{N_f g_s})^2)^{-1/2}$. Then if $E\approx \frac{1}{\lambda}$ that means $N_f g_s >>1$ and $g_s$ is strong. In this case the system should be described by Quantum Field Theory (QFT) in curved space where no perturbations show up. Hence our electric field is sending us to another theory such that our space is not flat any more. 

The effect of the electric field is clear in this work. $E$ increases the volume of D5-brane and decreases the low energy of the system and changes the nature of the background from flat to curved and tells us the system should now be studied in QFT in curved space.

We start the study by introducing D1$\bot$D5 branes and discussing the influence of the electric field on the low energy and the volume of D5-brane in section 2. We give the solutions of the linearized equations of motion of the relative transverse perturbations in flat space and we treat the effect of the electric field on the perturbations and the associated potentials in section 3. Then in section 4, we study the overall transverse perturbations and their associated potentials in zero and non-zero modes propagating on a dyonic string in the supergravity background of the orthogonal D5-branes and we look for the effect of the electric field in this case. The discussion and conclusion are presented in section 5.

\section{Intersecting D1 and D5 branes}
\hspace{.3in}Let's briefly review the non-abelian viewpoint of the ($N,N_f$)-strings which grow into D5-branes by using non-commutative coordinates \cite{dual,cm,supergravComp}. The dual picture is the intersecting D5 and D1 branes such that ($N,N_f$)-strings can end on D5-branes, but they must act as sources of second Chern class or instanton number in the world volume theory of the D5-branes. Hence D5 world volume description is complicated because of the second chern term which is not vanishing. The most important feature of the intersecting D1-D5 branes is the fact that the duality of this system discussed in \cite{d1d2d3d5duality} in the low energy effective theory with the electric field is switched on is unbroken.

In the present description, The fundamental $N_f$ strings are introduced by adding a $U(1)$ electric field denoted $F_{\tau\sigma}=EI_N$, with $I_N$ the $N\times N$ identity matrix. In fact the electric field turns the $N$ D-strings into a $(N,N_f)$-strings by dissolving the fundamental string degrees of freedom into the world volume.

For a fixed $E$ we consider the quantization condition on the displacement $D=\frac{N_f}{N}$ such that $D\equiv \frac{1}{N}\frac{\delta S}{\delta E}=\frac{\lambda^2 T_1 E}{\sqrt{1-\lambda^2 E^2}}$. Then the electric field is expressed in terms of string coupling $g_s$ and the number of fundamental strings $N_f$, \beq E=\frac{1}{\lambda}(1+(\frac{N}{N_f g_s})^2)^{-1/2}.\eeq

The electric field is turned on and the system dyonic is described by the action 
\beq S=-T_1\int d^2\sigma STr \Big[-det\pmatrix{\eta_{ab}+\lambda F_{ab}& \lambda \partial_a \Phi^j \cr -\lambda \partial_b \Phi^i & Q^{ij}\cr}\Big]^{1\over2},\eeq
with $i,j=1,...,5$, $a,b=\tau,\sigma$ and using $T_1 =\frac{1}{\lambda g_s}$ such that $\lambda=2\pi l_s^2$ with $l_s$ is the string length, $g_s$ is the string coupling and $Q_{ij}=\de_{ij}+i\lambda \lbrack \Phi_i , \Phi_j \rbrack$. The funnel solution is given by suggesting the ansatz 
\beq \Phi_i (\sigma)=\mp\hat{R}(\sigma)G_i,\eeq $i=1,...,5$, where $\hat{R}(\sigma)$ is the (positive) radial profile and $G_i$ are the matrices constructed  by Castelino, Lee and Taylor in \cite{clt}. We note that $G_i$ are given by the totally symmetric $n$-fold tensor product of 4$\times$4 Euclidean gamma matrices, such that $\frac{1}{2}[G^i , G^j]$ are generators of $SO(5)$ rotations, and that the dimension of the matrices is related to the integer $n$ by $N=\frac{(n+1)(n+2)(n+3)}{6}$. The Funnel solution (3) has the following physical radius \beq R(\sigma)=\sqrt{c}\lambda\hat{R}(\sigma),\eeq with $c$ is the Casimir associated with the $G_i$ matrices, given by $c=n(n+4)$, and the funnel solution is
\beq \Phi_i (\sigma)=\pm\frac{R(\sigma)}{\lambda\sqrt{c}}G_i.\eeq

We compute the determinant in (2) and we obtain
\beq S=-NT_1 \int d^2\sigma\sqrt{1-\lambda^2 E^2+(R')^2}(1+4\frac{R^4}{c\lambda^2}).\eeq
This result only captures the leading large N contribution at each order in the expansion of the square root. Using the action $(6)$, we can derive the lowest energy $\xi_{min}$ as the electric field is present and $E\in ]0,1/\lambda[$, (the low energy in the case of intersecting D1-D5 branes when the electric field is absent was discussed in \cite{cm}) $$\xi=NT_1 \int d\sigma\Big( \Big(\sqrt{1-\lambda^2 E^2}\mp R' \Big( \frac{8R^4}{c\lambda^2}+\frac{16R^8}{c^2\lambda^4}\Big)^{\frac{1}{2}} \Big)^2 +\Big( R'\pm \sqrt{1-\lambda^2 E^2}\Big( \frac{8R^4}{c\lambda^2}+\frac{16R^8}{c^2\lambda^4}\Big)^{\frac{1}{2}} \Big)^2\Big)^{\frac{1}{2}}$$ and
\beq \xi_{min} = NT_1 \sqrt{1-\lambda^2 E^2}\int \Big( 1+\frac{4R^4}{c\lambda^2}\Big)^2 d\sigma.\eeq such that \beq R'=\mp \sqrt{1-\lambda^2 E^2}\Big( \frac{8R^4}{c\lambda^2}+\frac{16R^8}{c^2\lambda^4}\Big)^{\frac{1}{2}}.\eeq
The lowest energy (7) can be rewritten in the following expression 
\beq \xi_{min} = N_f g_s T_1 \frac{1-\lambda^2 E^2}{\lambda E}\int_0 ^{\infty}d\sigma+\frac{6N}{c}T_5  \sqrt{1-\lambda^2 E^2}\int_0 ^{\infty}\Omega_4 R^4 dR +NT_1  \sqrt{1-\lambda^2 E^2}\int_0 ^{\infty}dR -\Delta \xi.\eeq In this equation, $T_5 =\frac{T_1}{(2\pi l_s)^4}$ and we can interpret the four terms as follows; the first term is the energy of $N_f$ strings and the second is the energy of $\frac{6N}{c} \approx n$ (for large $N$) D5-branes and the third is of N D-strings running out radially across D5-brane world volume and the last term is a binding energy 
\beq
\Delta \xi =2NT_1 \sqrt{1-\lambda^2 E^2}\int\limits_0^{\infty} du u^4 (1+\frac{1}{2u^4}-\sqrt{1+1\over u^4})
\approx 1.0102T_1 l_s N c^{1\over 4}\sqrt{1-\lambda^2 E^2}.\eeq
This equation shows that the lowest energy is gotten more lowest as the value of electric field is more important.

The equation (6) can be solved in dyoinc case by considering various limits. For small $R$, the physical radius of the fuzzy funnel solution (5) is found to be \beq R(\sigma)\approx\frac{\lambda\sqrt{c}}{2\sqrt{2}\sqrt{1-\lambda^2E^2}\sigma},\eeq and for large $R$ the solution is \beq R(\sigma)\approx(\frac{\lambda^2 c}{\sqrt{18}\sqrt{1-\lambda^2E^2}\sigma})^{\frac{1}{3}},\eeq with an upper bound on the electric field $E< \frac{1}{\lambda}$ for both cases.

According to equations (11) and (12), we remark that as the higher order terms in the BI action would effect a transition from the universal small R behavior to the “harmonic” expansion at large R ($\sigma$ goes to zero). The effect we get at this stage when the electric field is turned on is that $R$ is going up faster as $\sigma$ goes to zero once $E$ reaches approximately $\frac{1}{2\lambda}$ as shown in {\bf figure 1}, and we are on D5-brane. It looks like the electric field increases the velocity of the the transition from strings to D5-branes world volume. Also we remark that D5 brane got highest radius once $E$ close to its critical value.
\begin{center}
\includegraphics[width=3in,height=3in]{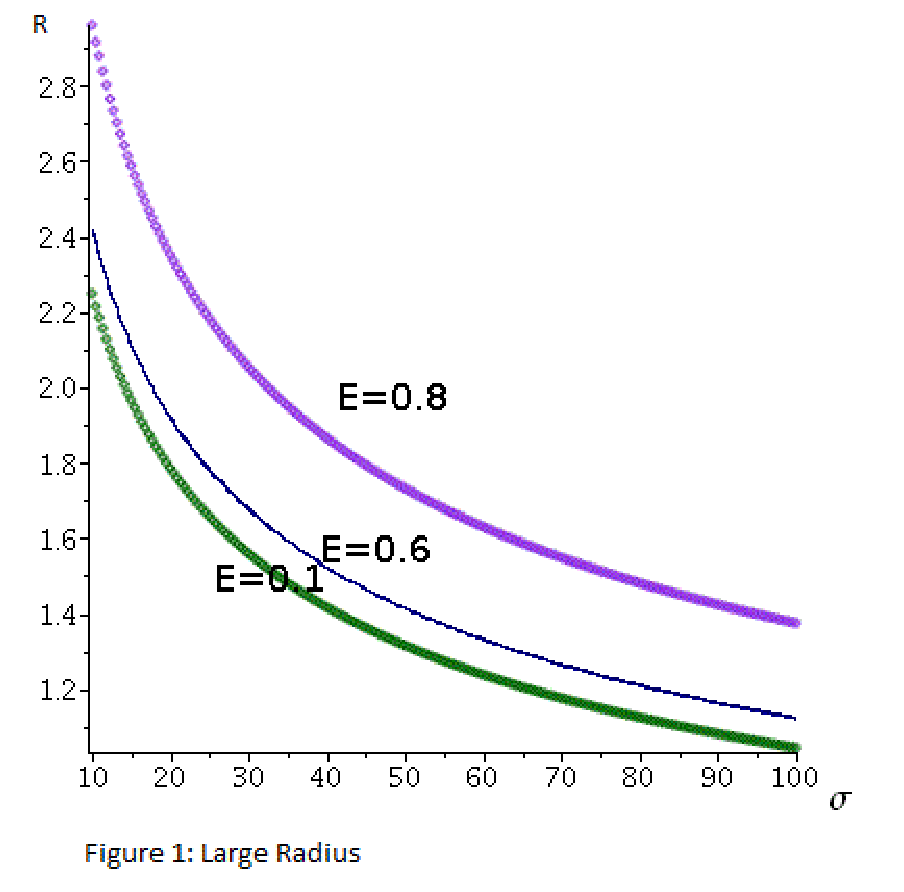}
\end{center}
The equations (9) and (12) give us the impression that the presence of the electric field is an important phenomena; it decreases the low energy and makes the D5-brane more voluminous.

In the following sections, we include a perturbation in the D5-brane configuration by simply adding lower and higher order symmetric polynomials in the $G^i$ to the Matrix configuration. We study the spatial perturbations of the moving D1-branes as the electric field is switched on.

\section{Flat Space}
\hspace{.3in}In this section, we examine the propagation of the perturbations on the fuzzy funnel by considering dyonic strings in flat background. We discuss the relative transverse perturbations which are transverse to the string, but parallel to the D5-brane world volume (i.e., along X$^{1,..,5}$). The overall transverse perturbations were studied in \cite{d1d3d5}.

We give the relative transverse perturbations in the following form \beq\delta \phi^i (\sigma,t)=f^i (\sigma,t)I_N,\eeq  as zero mode with  $i=1,..,5$ and $I_N$ the identity matrix. By inserting this perturbation into the full ($N,N_f$)-string action (2), together with the funnel (6) the action is found to be \beq S\approx-NT_1\int d^2\sigma  \Big[ (1-\lambda^2 E^2)A-(1-\lambda E)\frac{\lambda^2}{2}(\dot{f}^i)^2+\frac{(1+\lambda E)\lambda^2}{2A} (\partial_\sigma
f^i)^2 +...\Big],\eeq with \beq A=(1+\frac{4R(\sigma)^4}{c\lambda^2 })^2 .\eeq Then, in large and fixed $n$ the equations of motion are \beq\Big (\frac{1-\lambda E}{1+\lambda E}(1+\frac{n^2 \lambda^2}{16(1-\lambda^2 E^2)^2\sigma^4})^2\partial^{2}_{\tau}-\partial^{2}_{\sigma}\Big )f^i=0.\eeq

Let's suggest that $$f^i=\Phi(\sigma) e^{-iw\tau}\de x^i,$$ in the direction of $\de x^i$ with $\Phi$ is a function of $\sigma$ and the equations of motion become \beq\Big (-\frac{1-\lambda E}{1+\lambda E}(1+\frac{n^2 \lambda^2}{16(1-\lambda^2 E^2)^2\sigma^4})^2w^{2}-\partial^{2}_{\sigma}\Big )\Phi=0,\eeq which can be rewritten as \beq\Big (-\frac{1-\lambda E}{1+\lambda E}(\frac{n^2 \lambda^2}{8(1-\lambda^2 E^2)^2\sigma^4}+\frac{n^4 \lambda^4}{16^2(1-\lambda^2 E^2)^4\sigma^8})w^2-\partial^{2}_{\sigma}\Big )\Phi=\frac{1-\lambda E}{1+\lambda E}w^2\Phi.\eeq Since the equation looks complicated, we simplify the calculations by dealing with asymptotic analysis; we start by the system in small and then large $\sigma$ limits.
\subsection{Small $\sigma$ Region}
\hspace{.3in}In this region, we see that $\sigma^8$ dominates and the equation of motion is reduced to \beq\Big (-\partial^{2}_{\sigma}+V(\sigma)\Big )\Phi=\frac{1-\lambda E}{1+\lambda E}w^2\Phi,\eeq for each direction $\de x^i$, with the potential \beq V(\sigma)=-\frac{w^2n^4\lambda^4}{16^2(1+\lambda E)^5(1-\lambda E)^3\sigma^8}.\eeq The progress of this potential is shown in the figure 2; when we are close to the D5-brane the potential is close to zero and once $E$ is turned on it gets negative values until $E$ is close to its maxima, we see this potential goes down too fast to a very low amplitude minima ($-\infty$). This phenomena should have a physical meaning! This could be thought as a kink to increase the $\Phi$'s velocity to push the perturbation to disappear. 
\begin{center}
\includegraphics[width=3in,height=3in]{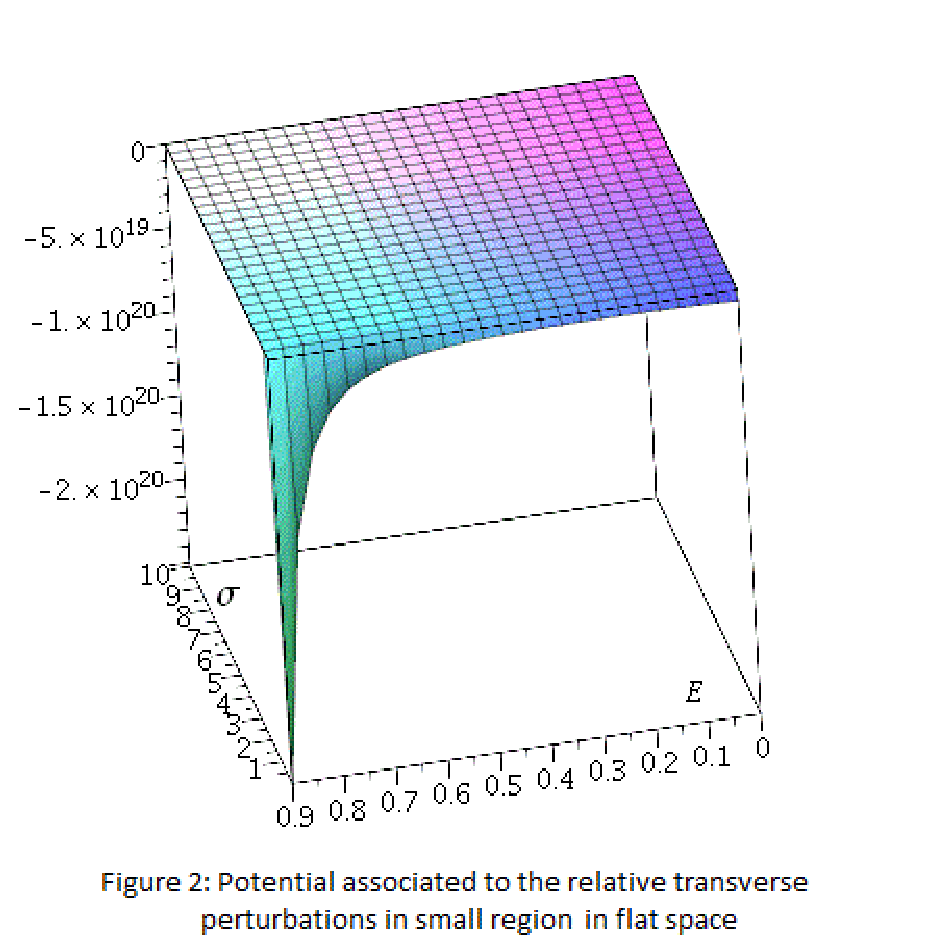}
\end{center}

To solve the equation (19), we consider the total differential on the perturbation. Let's denote $\partial_{\sigma}\Phi\equiv \Phi'$. Since $\Phi$ depends only on $\sigma$ we find $\frac{d\Phi}{d\sigma}=\partial_{\sigma}\Phi$. We rewrite the equation (19) in this form \beq \frac{1}{\Phi}\frac{d\Phi'}{d\sigma}=-w^2[\frac{n^4 \lambda^4 }{16^2(1+\lambda E)^5(1-\lambda E)^3\sigma^8}+1].\eeq An integral formula can be written as follows \beq \int\limits_{0}^{\Phi'}\frac{d\Phi'}{\Phi}=-\int\limits_{0}^{\sigma} w^2[\frac{n^4 \lambda^4 }{16^2(1+\lambda E)^5(1-\lambda E)^3\sigma^8}+1]d\sigma,\eeq which gives \beq \frac{\Phi'}{\Phi}=-w^2[-\frac{n^4 \lambda^4 }{16^2(1+\lambda E)^5(1-\lambda E)^3\times 7\sigma^7}+\sigma]+\alpha.\eeq We integrate again the following \beq\int\limits_{0}^{\Phi}\frac{d\Phi}{\Phi}=-\int\limits_{0}^{\sigma}(w^2[-\frac{n^4 \lambda^4 }{16^2\times 7(1+\lambda E)^5(1-\lambda E)^3\sigma^7}+\sigma]+\alpha) d\sigma.\eeq We get \beq\ln\Phi=-w^2[-\frac{n^4 \lambda^4 }{16^2\times 42(1+\lambda E)^5(1-\lambda E)^3\sigma^6}+\frac{2}{\sigma^2}]+\alpha\sigma+\beta,\eeq and the perturbation in small $\sigma$ region is found to be \beq \Phi (\sigma)=\beta e^{-w^2 [-\frac{n^4 \lambda^4 }{16^2\times 42(1+\lambda E)^5(1-\lambda E)^3\sigma^6}+\frac{\sigma^2}{2}]+\alpha\sigma},\eeq with $\beta$ and $\alpha$ are constants.

We plot the progress of the obtained perturbation. First we consider the constants $\beta=1=\alpha$, then the small spatial coordinate in the interval $[0,10]$ with the unit of $\lambda=1$, $w=1$ and $n\approx 10^3$ with the electric field in $[0,1[$.
\begin{center}
\includegraphics[width=3in,height=3in]{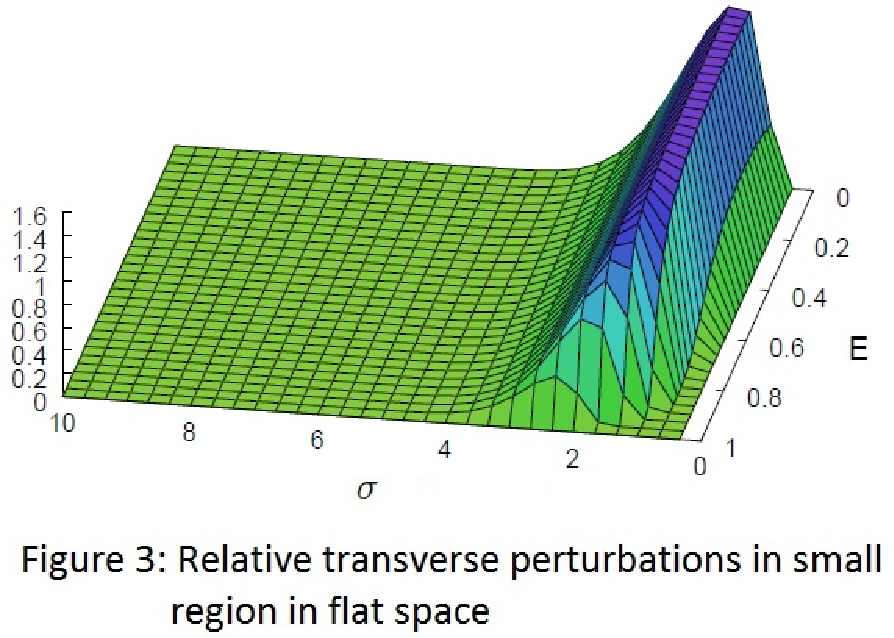}
\end{center}

As shown in figure 3, close to D5-brane there is perturbation. We remark that as $E$ goes up, the perturbation goes down. And when $E\approx 1/\lambda$ we observe no perturbation effects. At this stage, according to (1) the string coupling gets strong $N_f g_s >>1$ which means the system background is changed. We know that with strong coupling the system should be in supergravity background where the perturbations are no more. Consequently, the presence of $E$ kills the perturbation and moves the system from flat to supergravity background.

\subsection{Large $\sigma$ Region}
\hspace{.3in}By considering large $\sigma$ limit the equation of motion (18) becomes 
\beq\Big (-\partial^{2}_{\sigma}+V(\sigma)\Big )\Phi=\frac{1-\lambda E}{1+\lambda E}w^2\Phi,\eeq with the potential 
\beq V(\sigma)=-\frac{w^2n^2\lambda^2}{8(1+\lambda E)^3(1-\lambda E)\sigma^4}.\eeq By plotting the progress of this potential (figure 4) we remark that when $\sigma$ goes far a way from the D5 brane the potential vanishes approximately for all values of the electric field. And close to D5-brane the potential gets negative values. The effect of $E$ is very clear; as $E$ goes up $V$ slows down the decreasing until the medium of $E$, then $V$ decreases too fast until its minimum value for $E$ going up to its critical value. 

Consequently, the electric field has the same effect on $V$ in both regions of $\sigma$; as $E$ goes to its maxima $V$ goes to its minima.
 
\begin{center}
\includegraphics[width=3in,height=3in]{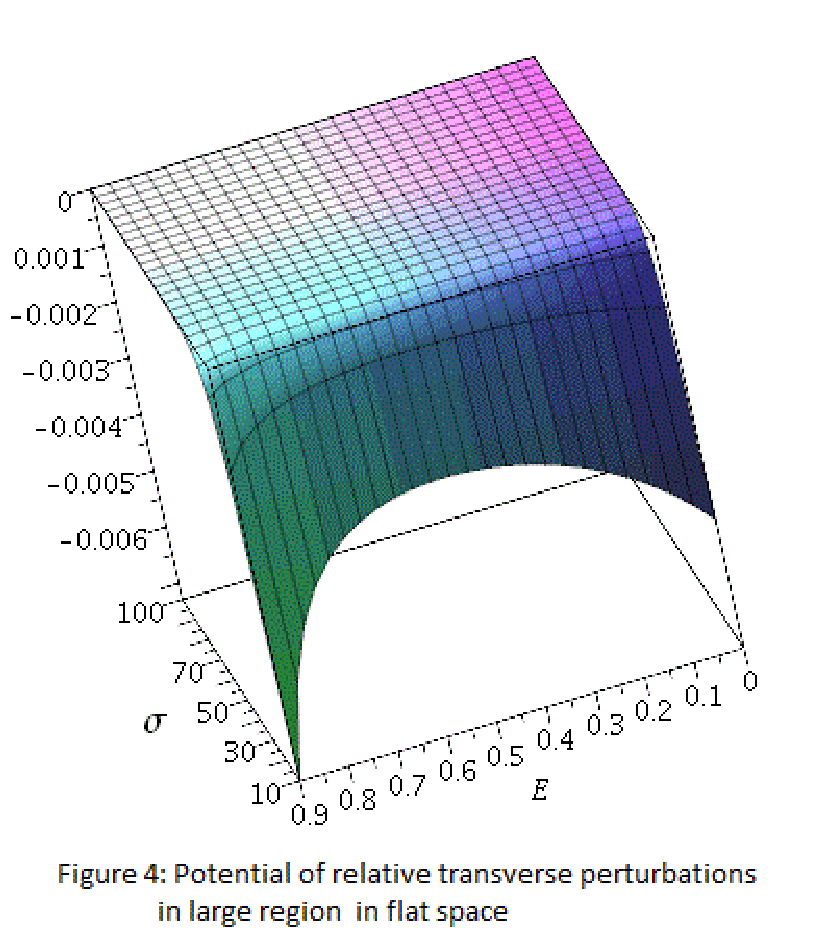}\includegraphics[width=3in,height=3in]{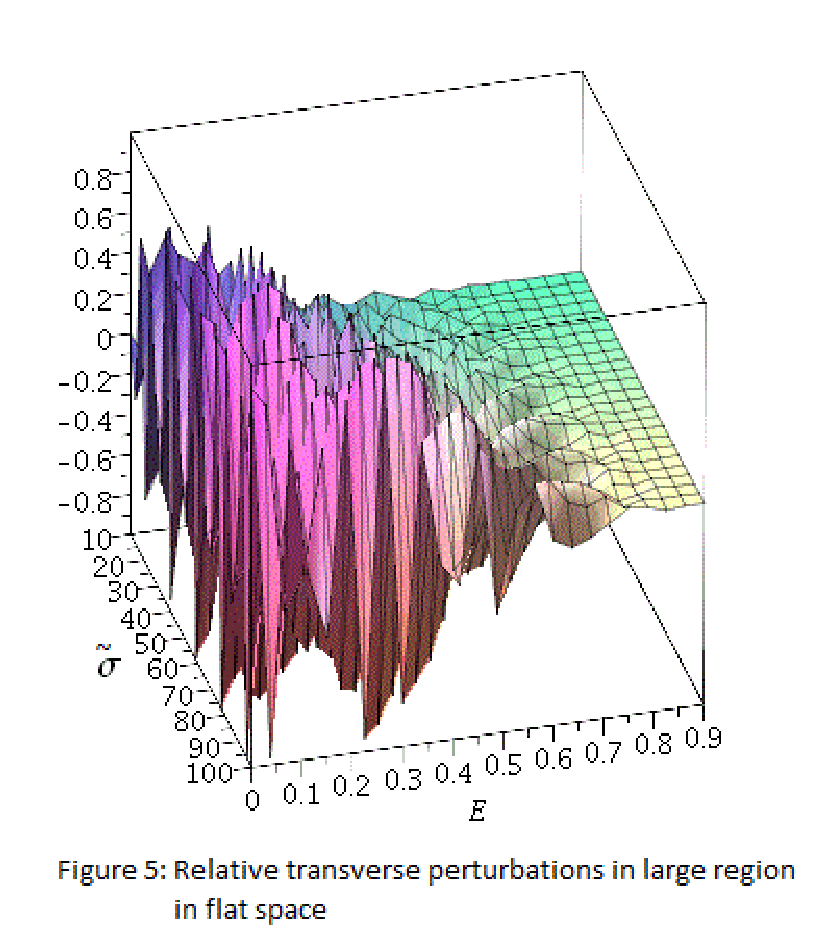}
\end{center}

To solve The equation (27) we rewrite it in the following form 
\beq\Big (\partial^{2}_{\tilde{\sigma}}+\frac{\kappa^2}{\tilde{\sigma}^4}+1\Big )\Phi=0,\eeq with 
\beq\tilde{\sigma}=\sqrt{\frac{1-\lambda E}{1+\lambda E}}w\sigma,\eeq and 
\beq\kappa^2=\frac{n^2\lambda^2}{8w^2(1+\lambda E)(1-\lambda E)^3},\eeq

The equation (29) is a Schr\"odinger equation for an attractive singular
potential $\propto\tilde{\sigma}^{-4}$ and depends on the single
coupling parameter $\kappa$ with constant positive Schr\"odinger
energy. The solution is then known by making the following
coordinate change 
\beq
\chi(\tilde{\sigma})=\int\limits^{\tilde{\sigma}}_{\sqrt{\kappa}}
dy\sqrt{1+\frac{\kappa^2}{y^4}}, \eeq and 
\beq
\Phi=(1+\frac{\kappa^2}{\tilde{\sigma}^4})^{-\frac{1}{4}}\tilde{\Phi}.
\eeq Thus, the equation (29) becomes \beq\Big(
-\partial^{2}_{\chi}+V(\chi)\Big) \tilde{\Phi}=0,\eeq with 
\beq
V(\chi)=\frac{5\kappa^2}{(\tilde{\sigma}^2
+\frac{\kappa^2}{\tilde{\sigma}^2})^3}.\eeq Then, the perturbation is found to be 
\beq
\Phi=(1+\frac{\kappa^2}{\tilde{\sigma}^4})^{-\frac{1}{4}}e^{\pm i\chi(\tilde{\sigma})}, \eeq which has the following limit; since we are in large $\sigma$ region $\Phi\sim e^{\pm i\chi(\tilde{\sigma})}$. This is the asymptotic wave function in the region $\chi\rightarrow +\infty$, while around $\chi\sim 0$, i.e. $\tilde{\sigma}\sim\sqrt{\kappa}$ and $\sigma\sim\frac{n\lambda}{2\sqrt{2}w^2 (1-\lambda E)^2}$, $\Phi\sim 2^{-\frac{1}{4}}$. 

Owing to the plotting of the progress of this perturbation (figure 5), by considering the real part of the function, the perturbation solution is totally different from the one gotten in small $\sigma$ limit (26). Hence the perturbations have a discontinuity and the system is divided into two regions which implies Neumann boundary conditions and the end of an open string can move freely on the brane in dyonic case, which means the end of a string on D5-brane can be seen as an electrically charged particle.

Figure 5 shows that the perturbation is slowing down as $E$ is turned on then starts to disappear once $E$ reaches the value $\frac{1}{2\lambda}$. The perturbation disappears when $E$ is too close to $\frac{1}{\lambda}$ for all values of $\sigma$. The effect of $E$ is very surprising! The presence of $E$ stops the perturbations.

No electric field means the intersecting point is in high perturbation. Then as $E$ is turned on the perturbations decrease. When $E$ is close to its critical value the perturbations are no more. They are killed by $E$. This phenomena matches very well with the fact that $g_s$ becomes strong ($N_f g_s >>1$) at this point according to the relation (5) such that $E\approx\frac{1}{\lambda}$. Consequently, we can suggest that the presence of the electric field changes the background of D-branes from flat to supergravity background (where the string coupling is strong).

\section{Curved Space}
\hspace{.3in}We extend the investigation of the intersecting D1-D5 branes to curved space. We consider again the presence of electric field and the resulting configuration is a bound state of fundamental strings and D-strings. Under these conditions the bosonic part of the effective action is the Non-Abelian BI action
\beq S=-T_1\int d^2\sigma e^{-\phi}STr \Big[-det\Big(P( G_{ab}+G_{ai}(Q^{-1}-\delta)^{ij}G_{jb}+\lambda F_{ab})\Big) det Q^{ij}\Big]^{1\over2},\eeq with $T_1$ the D1-brane tension, G the bulk metric, (for simplicity we set the Kalb-Ramond two form B to be zero), $\phi$ the dilaton and $F$ the field strength. $a,b=\tau,\sigma$ and $i,j=1,2,3,4,5$. Furthermore, $P$ denotes the pullback of the bulk space time tensors to each of the brane world volume. The matrix $Q$ is given by $Q^i_j =\delta^i_j +i\lambda [\phi^i ,\phi^k]G_{kj}$, with $\phi^i$ are the transverse coordinates to the D1-branes.

We consider the supergravity background and the metric of $n$ D5-branes
\beq\bra{lll}
ds^2 &=\frac{1}{\sqrt{h}} \eta_{\mu\nu}dx^{\mu} dx^{\nu}+\sqrt{h}(d\sigma^2 +\sigma^2 d\Omega_3^2)\\
e^{-\phi}&=\sqrt{h}\\
h&=1+\frac{L^2 }{\sigma^2},
\era\eeq
with $\mu , \nu=\tau , \sigma$ and $L=n l_s^2 g_s$.

\subsection{Zero Mode}
\hspace{.3in}In our work we treat $E$ as a variable to discuss its influence on the perturbations. We investigate the perturbations in the supergravity background of an orthogonal 5-brane in the context of dyonic stings growing into D5-branes. The study is focused on overall transverse perturbations in the {\bf zero mode}; $\delta\phi^i =f^i (\tau,\sigma)I$, $i=6,7,8,9$ and $I$ is $N\times N$ identity matrix.

The action describing the perturbed intersecting D1-D5 branes in the supergravity background is 
\beq\bra{ll} S &\equiv -NT_1 e^{-\phi}\int d^2\sigma \Big[ G_{\tau\tau}G_{\sigma\sigma}(1+\lambda E)-\frac{\lambda^2 }{2}(1-\lambda^2 E^2)G_{\sigma\sigma}G_{ii}(\dot{f}^i)^2 +\frac{\lambda^2}{2}(1+\lambda E)G_{\tau\tau}G_{ii}(f^i)'^2 +...\Big]\\
&\equiv -NT_1 \int d^2\sigma \sqrt{h}\Big[ 1+\lambda E-\frac{\lambda^2 \alpha_i}{2h}(1-\lambda^2 E^2)(\dot{f^i})^2 +\frac{\lambda^2 \sqrt{h}\alpha_i}{2}(1+\lambda E)(f^i)'^2 +...\Big],\era\eeq where $h(\sigma)=e^{-2\phi}=1+\frac{L^2}{\sigma^2}$, $\dot{f^i}=\partial_{\tau}f^i$, $(f^i)' =\partial_{\sigma}f^i$, $G_{\tau\tau}=h^{-1/2 }G_{\sigma\sigma}=\sqrt{h}e^{-\phi}$ and $G_{ii}=\alpha_i$ with $\alpha_i$ some real numbers.  

The equations of motion of the perturbations are found to be
\beq\Big( \frac{1-\lambda E}{h^{3/2}}\partial^2_\tau -\partial^2_\sigma+\frac{L^2}{h\sigma^3}\partial_\sigma\Big)f^i=0.\eeq If we consider $\tilde{\sigma}^2=\sigma^2+L^2$ the equations of motion become 
\beq\Big( \frac{1-\lambda E}{\sqrt{h}}\partial^2_\tau -\partial^2_{\tilde{\sigma}}\Big)f^i (\tilde{\sigma},t)=0.\eeq
We define the perturbations as \beq f^i (\tilde{\sigma},t)=\Psi(\tilde{\sigma})e^{-iw\tau}\delta x^i \eeq with $\delta x^i$ ($i=6,7,8,9$) the direction of the perturbation and the equation (41) becomes \beq (-w^2 (1-\lambda E)\frac{\tilde{\sigma}}{\sqrt{\tilde{\sigma}^2-L^2}}-\partial^2_{\tilde{\sigma}})\Psi=w^2 (1-\lambda E)\Psi\eeq with the potential $V=-w^2 (1-\lambda E)\frac{\tilde{\sigma}}{\sqrt{\tilde{\sigma}^2-L^2}}=-w^2 (1-\lambda E)\frac{\sqrt{\sigma^2+L^2}}{\sigma}$.
\begin{center}
\includegraphics[width=3in,height=3in]{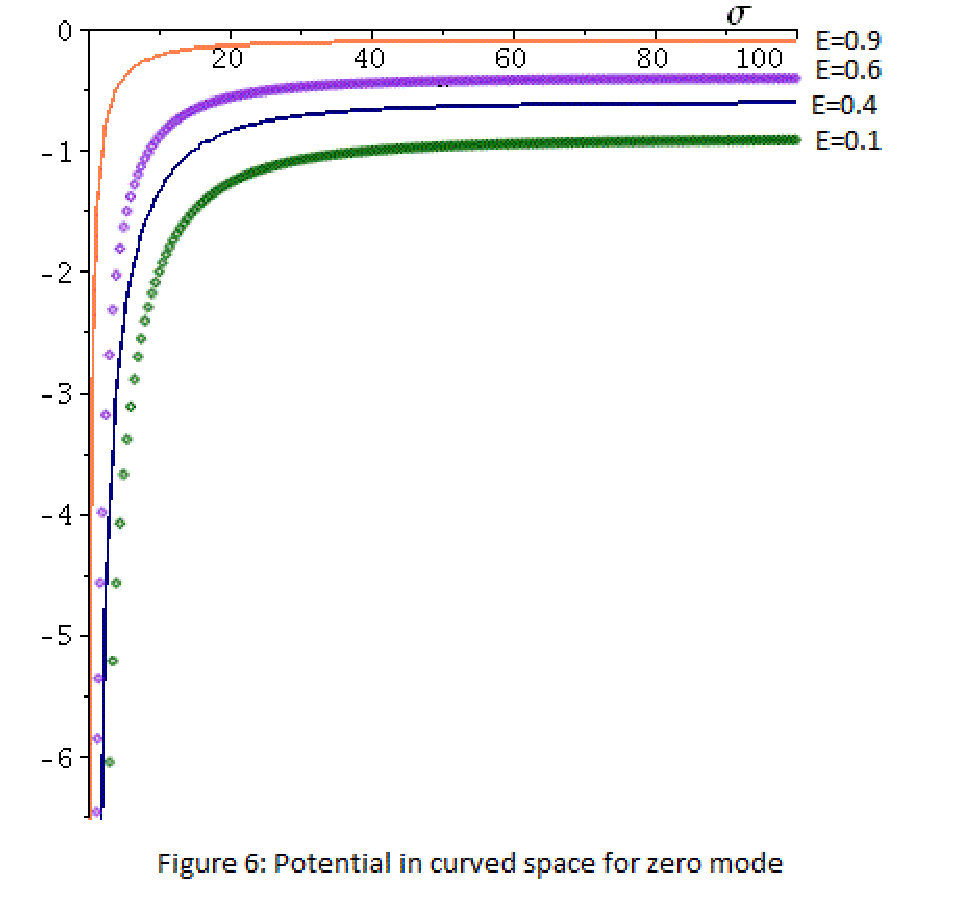}\includegraphics[width=3in,height=3in]{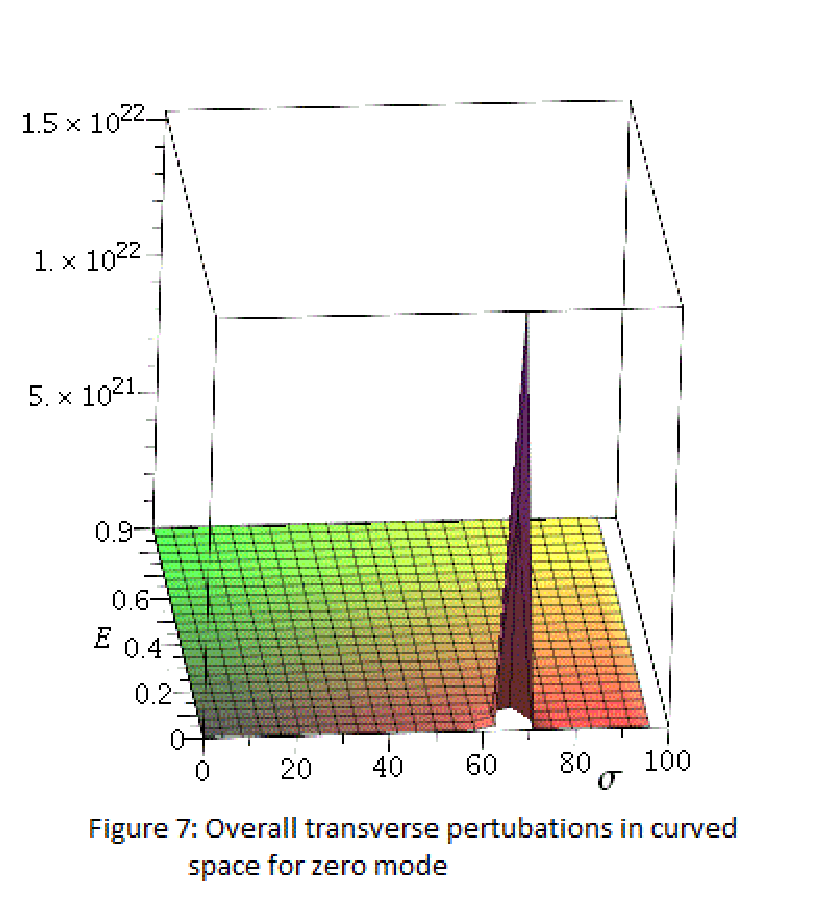}
\end{center}
The figure 6 shows the variation of the potential $V$ in terms of $\sigma$. We remark approximately the absence of the potential for all large values of $\sigma$ and $V$ goes to zero as $E$ goes to $\frac{1}{\lambda}$. When $\sigma$ is too close to zero, in this case $V$ is negative and goes down too quick for all $E$ and the potential is not that low. In addition, in the curved space the effect of $E$ approximately absent.

Let's solve the differential equation (43). As we see this is the Heun’s equation and the solution is the perturbation
\beq\bra{ll}\Psi &=(-\tilde{\sigma}^2 +L^2)\Big[ \eta HeunC(0,\frac{-1}{2},1,\frac{1}{4} w^2(1-\lambda E)L^2 ,\frac{1}{2} +\frac{1}{4} (-L^2 +L^2)w^2(1-\lambda E),\tilde{\sigma}^2 /L^2)\\&+\beta HeunC(0,\frac{1}{2},1,\frac{1}{4} w^2(1-\lambda E)L^2 ,\frac{1}{2} +\frac{1}{4} (-L^2 +L^2)w^2(1-\lambda E),\tilde{\sigma}^2 /L^2)\Big]\tilde{\sigma}\era\eeq with $\eta$ and $\beta$ are constants.

We tried to plot the perturbation (44) for small region of $\sigma$ (the radius of funnel solution is too large) and there is no perturbation in this region. The intersecting point is stable in supergravity background even if the electric field is present.

The figure 7 shows the variation of the perturbation in terms of the electric field $E$ and the coordinate $\tilde{\sigma}$ in large region such that the radius of funnel solution is too small. We set $\lambda=1$, $w=1$ and $n=10^2$. The perturbation is showing up as a peak for a while and for low electric field. In general we observe approximately no perturbation effects for all $E$ in this case.

The important remark we obtain by comparing the influence of $E$ on the perturbation in flat and curved spaces is that $E$ kills the perturbation in flat space (Fig.3, Fig.5) and turns the string coupling to be strong and then the flat space in this case becomes curved when $E$ reaches its critical value, but when the space is already curved the influence of $E$ is absent. This observation leads us to think that $E$ is strongly related in some way to the supergravity background. 

\subsection{Non-Zero Modes}
Let's now consider the {\bf non-zero modes} and the perturbations could be written in this form $\delta \phi^m (\sigma,t)=\sum\limits^{N-1}_{\ell=1}\psi^{m}_{i_1 ...i_\ell}G^{i_1} ... G^{i_\ell} $ with $\psi^{m}_{i_1 ... i_\ell}$ are completely symmetric and traceless in the lower indices. We get two terms added to the action (39) to describe the present system $\lbrack \phi^i ,\delta\phi^m \rbrack^2 $ and $\lbrack \partial_{\sigma}\phi^i ,\partial_{t}\delta\phi^m \rbrack^2 $. Then in the equation of motion (40) these two terms appeared $\lbrack \phi^i ,\lbrack \phi^i ,\delta\phi^m \rbrack\rbrack$ and  $\lbrack \partial_{\sigma}\phi^i ,\lbrack\partial_{\sigma}\phi^i ,\partial^2 _{t}\delta\phi^m \rbrack\rbrack$. We have $\phi^i =RG^i$ and by straightforward calculations we have
\beq\bra{lll} \lbrack G^i ,\lbrack G^i, \delta\phi^m \rbrack\rbrack
&=\sum\limits_{\ell<N}^{N-1}\psi^{m}_{i_1 ... i_\ell}\lbrack G^i ,\lbrack G^i ,G^{i_1} ... G^{i_\ell} \rbrack\rbrack\\\\
&=\sum\limits_{\ell<N}^{N-1}\psi^{m}_{i_1 ...i_\ell}\epsilon^{i_1 ...i_{\ell}}G^{i_1} ... G^{i_\ell},\\\\
&=\sum\limits_{\ell<N}^{N-1}4\ell(\ell+\beta)\delta\phi^m_\ell
\era\eeq with $\epsilon^{i_1 ...i_{\ell}}$ antisymmetric tensor and $\beta$ a real number. To obtain a specific spherical harmonic on 4-sphere, we have 
\beq\lbrack \phi^i ,\lbrack \phi^i ,\delta\phi_{\ell}^m \rbrack\rbrack=\frac{\ell(\ell+\beta) \lambda^2 c}{2(1-\lambda^2E^2)\sigma^2}\delta\phi_{\ell}^m ,\phantom{~~~~~~}\lbrack \partial_{\sigma}\phi^i ,\lbrack\partial_{\sigma}\phi^i ,\partial_{t}^2 \delta\phi^m \rbrack\rbrack=\frac{\ell(\ell+\beta) \lambda^2 c}{2(1-\lambda^2E^2)\sigma^4}\partial_{t}^2\delta\phi_{\ell}^m .\eeq Then for each mode we set $\delta\phi^m_\ell =f^m_\ell (\tilde\sigma)e^{-i\omega\tau}\delta x^m$ with $f^m_\ell$ some function for each mode. Then the equations of motion will be in this form \beq (-\partial^2_{\tilde\sigma}+V(\tilde\sigma))f^m_\ell (\tilde\sigma)=-w^2 (1-\lambda E)f^m_\ell (\tilde\sigma),\eeq with $V(\tilde\sigma)=V_1 +V_2 +V_3$ and \beq V_1 =-w^2 (1-\lambda E)\frac{\tilde\sigma}{\sqrt{\tilde\sigma^2 -L^2}}=-w^2 (1-\lambda E)\frac{\sqrt{\sigma^2 +L^2}}{\sigma},\eeq\\
\beq V_2 =\frac{\ell(\ell+\beta)\lambda^2 c}{2(\tilde\sigma^2 -L^2)}=\frac{\ell(\ell+\beta)\lambda^2 c}{2\sigma^2 },\eeq\\
\beq V_3 =\frac{\ell(\ell+\beta)\lambda^6 cw^2 \al^i \al^m}{24(1-\lambda^2 E^2)(\tilde\sigma^2 -L^2)^2}=\frac{\ell(\ell+\beta)\lambda^6 cw^2 \al^i \al^m}{24(1-\lambda^2 E^2)\sigma^4}.\eeq These expressions can be treated by taking into account the limits of $\sigma$ such as $\sigma$ goes to zero and the infinity.

For small $\sigma$, $V_3$ dominates and in large $\sigma$, $V_1 +V_2$ will dominates. It's clear for now on that the system in the present background will get different potentials and perturbations from region to other which support the idea of Neumann boundary condition in supergravity background.

We start by small $\sigma$ region, and the plot of $V_3$ (figure 8) shows that if $\sigma$ goes to zero then the potential goes to $+\infty$. Physically this behavior should means something! This could be a sign to the absence of the perturbation effects and the influence of $E$ is absent.
\begin{center}
\includegraphics[width=3in,height=3in]{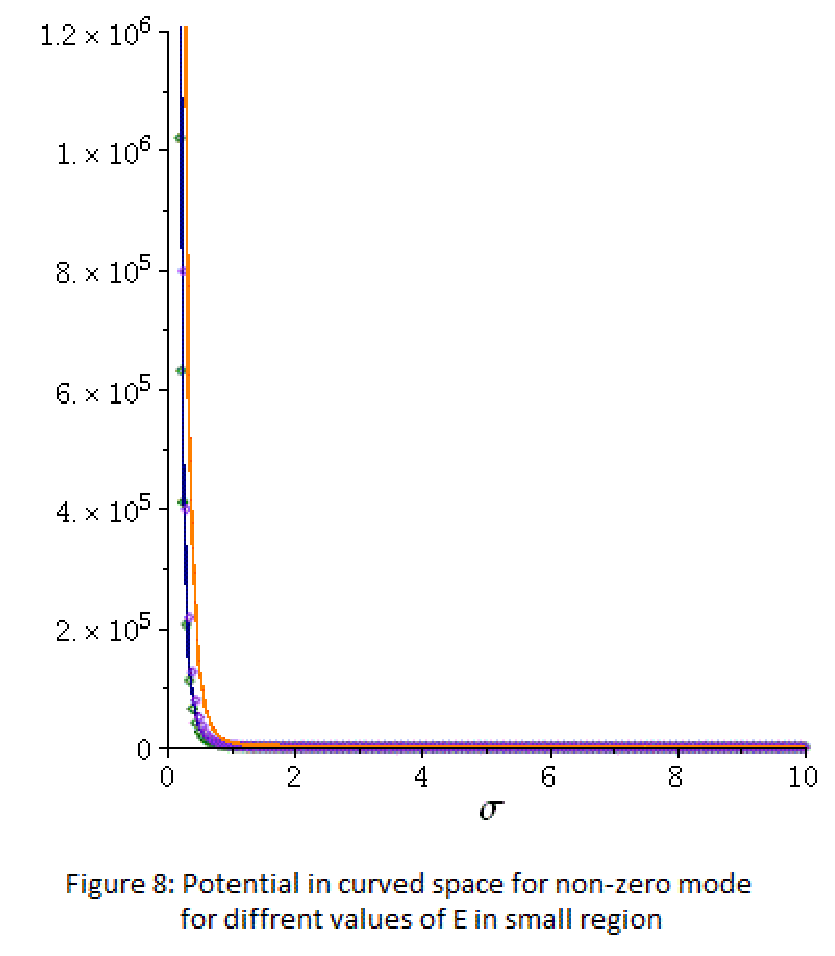}
\end{center}
We remark that the electric field does not have any influence on the perturbations in non-zero mode at the presence of the supergravity background.

Then the perturbation for each mode $\ell$ is gotten
\beq\bra{llll}
f^m_\ell = b_1 e^{-\frac{1}{24}\lambda^3 x \frac{(\frac{2}{3}\sqrt{-6 d (-1+\lambda^2 E^2)} x^2 (-d(-1+\lambda^2 E^2))^{2/3}+dL^2 2^(2/3) 12^{1/3} (-6 (-1+\lambda^2 E^2))^{1/6} (-1+\lambda E)(\lambda E+1))}{(-d (-1+\lambda^2 E^2))^{2/3}}}\\\\ HeunT\Big( \frac{-3 2^{1/3} d (-1+\lambda^2 E^2) (-1+E)}{\lambda^2 (-d (-1+\lambda^2 E^2))^{4/3}}, 0, \frac{\frac{1}{2} d \lambda^2 L^2 (-1+\lambda^2 E^2) 2^{2/3}}{(-d (-1+\lambda^2 E^2))^{2/3}}, \frac{12^{1/3} (-6 d (-1+\lambda^2 E^2))^{1/6} \lambda x}{6} \Big)\\\\+b_2 HeunT\Big( \frac{-3 2^{1/3} d (-1+\lambda^2 E^2) (-1+\lambda E)}{\lambda^2 (-d(-1+\lambda^2 E^2))^{4/3}}, 0, \frac{\frac{1}{2} d \lambda^2 L^2 (-1+\lambda^2 E^2) 2^{2/3}}{-d (-1+\lambda^2 E^2)^{2/3}}, -\frac{12^{1/3} (-6 d(-1+\lambda^2 E^2))^{1/6}\lambda x}{6} \Big)\\\\ e^{\frac{\frac{1}{24} \lambda^3 x (\frac{2}{3}\sqrt{-6d(-1+\lambda^2 E^2)} x^2 (-d (-1+\lambda^2 E^2))^{2/3}+d L^2 2^(2/3) 12^{1/3} (-6 d (-1+\lambda^2 E^2))^{1/6} (-1+\lambda E) (\lambda E+1))}{(-d (-1+\lambda^2 E^2))^{2/3}}}\era\eeq
with $b_1$ and $b_2$ are constants and $d =\ell(\ell +\beta)\lambda^6 n(n+1)\al^i \al^m w^2$. We tried to plot this function but noway we could not get any perturbation for the values $\lambda=1$, $w=1$ and for all $E$, $\ell>4$ and $n>1$ in the region $\sigma\in [0,10]$.

Let's move to the large $\sigma$. As $\sigma$ goes to infinity we see the potential goes to zero (figure 9) but when $\sigma$ approaches the small $\sigma$ region the potential goes up too quick and reaches the maximum value, approximately for all $E$. Then the electric field does not have influence on the behavior of the potential in curved space.
\begin{center}
\includegraphics[width=3in,height=3in]{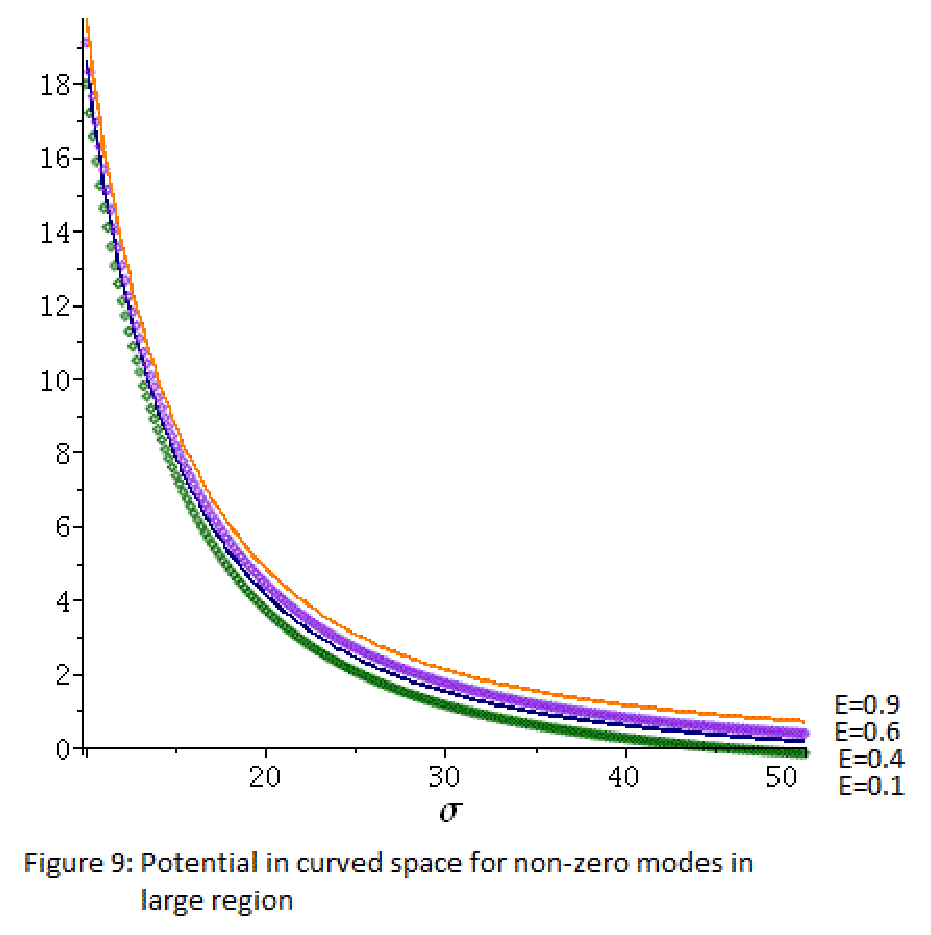}
\end{center}
Also the potential shows up with little values by comparison to the case  of small region and for all $E$ which means $E$ does not change anything in the case of curved space.

The perturbation for each mode is \beq\bra{llll} f^m_\ell = a_1 HeunC \Big( 0, \frac{\sqrt{2w^2 L^4 (\lambda E-1)+L^2 -4 \lambda^2 cl(l+\beta)}}{2L}, -2, \frac{w^2 L^2 (\lambda E-1)}{8}, \frac{5}{4}-\frac{w^2 L^2 (\lambda E-1)}{8}, \frac{2\tilde\sigma^2 -2\tilde\sigma\sqrt{\tilde\sigma^2 -L^2}-L^2}{L^2}\Big)\\
(\sqrt{\tilde\sigma^2 -L^2} +\tilde\sigma )^{\frac{L-\sqrt{2w^2 L^4 (\lambda E-1) + L^2 -4\lambda^2 cl(l+\beta)}}{2L}}\\+a_2 HeunC \Big( 0, -\frac{\sqrt{2w^2 L^4 (\lambda E-1)+L^2-4 \lambda^2 c l (l+\beta)}}{2L}, -2, \frac{w^2 L^2 (\lambda E-1)}{8}, \frac{5}{4}-\frac{w^2 L^2 (\lambda E-1)}{8}, \frac{2\tilde\sigma^2-2\tilde\sigma\sqrt{\tilde\sigma^2-L^2}-L^2}{L^2} \Big)\\(\sqrt{\tilde\sigma^2 -L^2}+\tilde\sigma)^{\frac{L+\sqrt{2w^2 L^4 (\lambda E-1)+L^2- 4\lambda^2 cl(l+\beta)}}{2L}},\era\eeq
with $a_1$ and $a_2$ are real constants. We tried to plot this function for all $E$, $\ell=10$ and $n=10^2$, and no perturbations appear which is consistent with the nature of space. Since the system is in supergravity background, there is no perturbations then no influence of electric field. 

\section{Discussion and Conclusion}
\hspace{.3in}In the low energy effective theory with the electric field $E$ is switched on, we proved in \cite{d1d3duality} that the duality of intersecting D1-D3 branes is broken and in \cite{d1d2d3d5duality} the duality of intersecting D1-D5 branes is unbroken. Hence, it is interesting to know more about the effect of the electric field, and the intersecting D1-D5 branes looks more important as a system.

We consider the non-abelian Born-Infeld (BI) dynamics of the dyonic string such that the electric field E has a limited value. If we suppose there is no excitation on transverse directions then the action of D1-branes is $S=-NT_1 \int d^2\sigma\sqrt{1-\lambda^2 E^2}.$ The limit of $E$ attains a maximum value $E_{max}=\frac{1}{\lambda}$ just as there is an upper limit for the velocity in special relativity. In fact, if $E$ is constant, after T-duality along the direction of $E$ the speed of the brane is precisely $\lambda E$ so that the upper limit on the electric field follows from the upper limit on the velocity. Hence if this critical value arises such as $E_{max}>\frac{1}{\lambda}$ the action ceases to make physical sense and the system becomes unstable. Since The string effectively carries electric charges of equal sign at each of its endpoints, as E increases the charges start to repel each other and stretch the string. For $E$ larger than the critical value, the string tension $T_1$ can no longer hold the strings together. 

In this context, we have treated in this project in particular the perturbations of a set of $(N,N_f)$-strings ending on a collection of $n$ orthogonal D5-branes in lowest energy world volume theory. The fundamental strings ending on an orthogonal D5-branes act as an electric point sources in the world volume theory of D5-brane and the perturbations in both flat and curved spaces were studied from this point of view. 

We showed in section 2 that the semi-infinite fuzzy funnel is a minimum energy configuration by imposing singular boundary conditions that have interesting physical interpretation in terms of D-brane geometries. And to consider the lowest energy effective theory the electric field should be present. We found the lowest energy $$\bra{llll} \xi_{min} &= N_f g_s T_1 \frac{1-\lambda^2 E^2}{\lambda E}\int_0 ^{\infty}d\sigma \\&+\frac{6N}{c}T_5  \sqrt{1-\lambda^2 E^2}\int_0 ^{\infty}\Omega_4 R^4 dR \\&+NT_1  \sqrt{1-\lambda^2 E^2}\int_0 ^{\infty}dR \\&-1.0102T_1 l_s N c^{1\over 4}\sqrt{1-\lambda^2 E^2}\era$$ by considering $E$ switched on in the low energy effective theory. The energy of intersecting D1-D5 branes is found to be a sum of four parts depending on the electric field $E$ and all these energies are decreasing as $E$ goes to $\frac{1}{\lambda}$. The first is for $N_f$ fundamental strings extending orthogonally away from the D5-branes and the second for the $n$ D5-branes and the third for the $N$ D-strings extending out radially in D5-branes and the forth is the binding energy.

In this theory, the transition between the universal behavior at small radius of the funnel solution and the harmonic behavior at large one in terms of electric field is mentioned too. When the electric field is turned on the physical radius of the fuzzy funnel solution $R(\sigma)\approx(\frac{\lambda^2 c}{\sqrt{18}\sqrt{1-\lambda^2E^2}\sigma})^{\frac{1}{3}}$ is going up faster as $\sigma$ goes to zero (the intersecting point) and $E$ reaches approximately $\frac{1}{2\lambda}$ which looks like the electric field increases the velocity of the transition from strings to D5-branes world volume. Then D5-branes get highest radius once $E$ is close to $\frac{1}{\lambda}$ which interprets the increasing of the volume of the D5-branes under the effect of the electric field (Figure 1).

In section 3, we have investigated the relative transverse perturbations of the funnel solutions of the intersecting D1-D5 branes in flat space and the associated potentials in terms of the electric field $E\in ]0,1/\lambda[$ and the spatial coordinate $\sigma$. We find that too close to the intersecting point the potential is close to zero and once E is turned on it gets negative values until $E$ is close to its maxima, we see this potential goes down too fast to a very low amplitude minima $-\infty$ (figures 2,4) and away from the intersecting point there is approximately no potential for all $E$. This is interpreted as inducing an increase in the velocity of the perturbation to disappear at the intersecting point toward the D5-brane world volume. Figures 3,5 show that when $E$ goes to its maxima there is no perturbation effects. Hence the presence of E kills in general the perturbations. At this stage, according to (1) the string coupling starts to get strong which means the system background is changing.

In curved space, we have studied the same system by looking for the effect of electric field on the perturbations and the associated potentials in zero (figures 6,7) and non zero-modes (figures 8,9) of the overall transverse perturbations in section 4. It was surprisingly that too close to the intersecting point; i.e. at large physical radius of D5-brane, we could not find any perturbation and also there is approximately no influence of $E$ on potentials. The effect of $E$ appears only when we are too far away from the intersecting point where the radius is too small and still $E$ makes the perturbations to disappear on the strings. In general we do not see the influence of $E$ in curved space.

The main and very important feature we got from this investigation is the following; the presence of electric field flux on the strings changes the background of the system. We proved explicitly that when the coupling is going to be strong which means $E$ goes to its critical value we should move to QFT to describe the system where no perturbations exist. In curved space the influence of the electric field appears for too small radius of funnel solution which means for large spatial coordinate $\sigma$ of strings and this phenomena decreases from zero mode to non-zero modes but when the radius is important as $\sigma$ goes to zero there is no effect of $E$. By contrast in the case of flat space that was very clear when $E$ is turned on the perturbations change their behavior in general. $E$ forces them to disappear as it is close to the critical value and in meantime the string coupling is getting strong.

The string coupling is strong means $N_f g_s >>1$ and  $g_s \approx \frac{N}{N_f}$ since $E\approx \frac{1}{\lambda}$ which is the critical value and if the electric field exceeds this value the system will be non-physical phenomena as discussed above and to be out of this problem we should choose another theory to describe our system.

In the case of weak coupling $N_f g_s <<1$ the electric field will be approximately $E\approx \frac{N_f g_s}{\lambda N}$ and the condition matches our perturbative phenomena $E\in [0,\frac{1}{\lambda}[$ . We mention here that if $E$ goes to zero then $ N_f g_s$ does too which means the number of fundamental strings decreases and simply the endpoints of the strings loose their electric charges and vice-versa.

In curved space, we can say the electric field $E$ has no effect on the intersecting point. We can connect then the phenomena to the electric field $E$ and the string coupling $g_s$ such as $E$ and $g_s$ are connected by the relation (5). We see that once $E$ is turned on and goes up $g_s$ is getting stronger. At the critical point, $E$ reaches its maxima and $g_s$ is strong then the space should become curved. Hence we can remark at this stage that the effect of $E$ looks like it transforms the flat space to curved one. In this context we can say there is a one-to-one map between the supergravity background and the electric field that we should look for!

\section*{Acknowledgments}
\hspace{.3in}The author would like to thank the Abdus Salam International Centre for Theoretical Physics, Trieste, for the invitation and the hospitality during the stage in which this paper was done.


\end{document}